\documentclass[11pt]{article}

\usepackage[a4paper,margin=1in]{geometry}
\usepackage{graphicx}
\usepackage{amsmath,amssymb,bm}
\usepackage{booktabs}
\usepackage{siunitx}
\usepackage[numbers,sort&compress]{natbib}
\usepackage{hyperref}
\usepackage{caption}
\usepackage{subcaption}
\usepackage{microtype}

\hypersetup{
  colorlinks=true,
  linkcolor=blue,
  citecolor=blue,
  urlcolor=blue
}
\title{\textbf{sidmkit: A Reproducible Toolkit for SIDM Phenomenology and Galaxy Rotation-Curve Modeling}}

\author{
  Nalin Dhiman \\[4pt]
  \small School of Computing and Electrical Engineering \\
  \small Indian Institute of Technology Mandi, India \\
  \small \texttt{d24008@students.iitmandi.ac.in}
}

\date{\today}

\begin{document}
\maketitle

\begin{abstract}
Self-interacting dark matter (SIDM) is a well-motivated extension of cold dark matter that can modify halo structure on galactic and group scales while remaining consistent with large-scale structure. However, practical SIDM work often requires bridging several layers, including microphysical scattering models, velocity-dependent effective cross sections, phenomenological astrophysical constraints, and (separately) data-driven halo fits, such as rotation curves. In this paper, we describe \texttt{sidmkit}, a transparent and reproducible Python package designed to support SIDM ``micro$\rightarrow$macro'' calculations and to provide a robust batch pipeline for fitting rotation curves in the SPARC data. On the SIDM side, \texttt{sidmkit} implements velocity-dependent momentum-transfer cross sections for a Yukawa interaction using standard analytic approximations (Born, classical, and Hulth\'en-based) with a numerical partial-wave option for spot checks. It also provides consistent velocity-moment averaging for Maxwellian relative speeds, scattering-rate utilities, and curated literature \emph{summary} constraints for regression tests and exploratory scans. On the rotation-curve side, we implement bounded non-linear least squares fits of NFW and Burkert halo models to SPARC baryonic decompositions, with optional mass-to-light priors and information-criterion summaries (AIC/BIC). For the demonstration dataset, we process 191 \texttt{rotmod} galaxies (LTG+ETG bundles) and fit both NFW and Burkert models (382 total fits). We find that Burkert is preferred by $\Delta \mathrm{BIC} > 0$ for $65.4\%$ of galaxies, with ``strong'' preference ($\Delta \mathrm{BIC}>6$) in $32.5\%$ of galaxies; NFW is strongly preferred in $14.7\%$. Median reduced $\chi^2$ values are $1.25$ (NFW) and $0.71$ (Burkert) for cases with positive degrees of freedom. These results summarise phenomenological fit quality and should \emph{not} be interpreted as a direct SIDM measurement without a careful treatment of baryonic, geometric, and selection systematics.We stress reproducibility and honesty with reviewers: the package is meant to be a reliable starting point, not a claim of definitive astrophysical inference. The toolkit is an open-source Python package that the community can use to do analyses and add to it.
\end{abstract}

\section{Introduction}
The standard cold dark matter (CDM) paradigm has been remarkably successful on large scales; however, on galactic scales, several long-discussed tensions, such as the cusp-core problem and the diversity of rotation curves, motivate a careful scrutiny of dark matter microphysics and baryonic modelling. One minimal extension is self-interacting dark matter (SIDM), where dark matter particles scatter elastically with a cross section per unit mass $\sigma/m$ that can be velocity dependent \citep{SpergelSteinhardt2000, TulinYu2018, Adhikari2022Review}. In many SIDM models, scattering is efficient in dwarf and low-surface-brightness galaxies but suppressed in clusters, potentially producing cored density profiles in some systems while remaining consistent with cluster bounds \citep{Rocha2013SIDM, Kaplinghat2016Colliders}.

Turning SIDM from an idea into a quantitative analysis is, in practice, a workflow problem. A typical study requires:
(i) a microphysical model (e.g., Yukawa-mediated scattering),
(ii) a mapping from particle parameters to velocity-dependent effective cross sections,
(iii) astrophysical observables that depend on velocity moments (not just $\sigma/m$ at a single $v$),
(iv) halo-level quantities such as scattering rates and core formation radii,
and (v) independent empirical constraints or likelihoods from clusters, dwarfs, and rotation curves.
Different papers often implement overlapping pieces with slightly different conventions, units, and numerical approximations, making reproduction and cross-checking difficult.

In parallel, rotation-curve datasets such as SPARC \citep{Lelli2016SPARC} provide high-quality measurements and baryonic decompositions that are invaluable for testing halo phenomenology. Even when one is not performing a direct SIDM microphysical inference, robust batch fitting of standard halo profiles (cuspy and cored) is a useful calibration and sanity check layer.

This work introduces \texttt{sidmkit}, a small but rigorous toolkit aimed at \emph{transparent} SIDM micro$\rightarrow$macro calculations and \emph{reproducible} batch rotation-curve fits. The goals are:
\begin{itemize}
\item \textbf{Correctness-first}: stable units, explicit assumptions, and built-in regression benchmarks;
\item \textbf{Modularity}: microphysics, averaging, constraints, halo utilities, and SPARC fitting are separated but composable;
\item \textbf{Honest scope}: provide baseline tools and summaries without overstating inference.
\end{itemize}

\section{Software overview}
\texttt{sidmkit} is distributed as a standard Python package with both an importable API and a command-line interface (CLI). The design philosophy is to keep the public API small and to make most workflows runnable from the CLI for reproducibility. The main components are:
\begin{enumerate}
\item \textbf{Microphysics layer:} Yukawa model definition and $\sigma_T(v)$ computation.
\item \textbf{Velocity averaging:} numerical evaluation of $\langle \sigma v^n\rangle/m$ for Maxwellian relative speeds.
\item \textbf{Constraint layer:} curated \emph{summary} constraints and simple point-likelihood helpers.
\item \textbf{Halo utilities:} scattering rate estimates and an illustrative $r_1$ core-formation radius calculation.
\item \textbf{SPARC batch fitter:} NFW and Burkert fits to \texttt{rotmod} files with chunking (\texttt{--skip}, \texttt{--limit}) and merged population reports.
\end{enumerate}

The package relies primarily on \texttt{NumPy} and \texttt{SciPy} for numerics \citep{Harris2020NumPy, Virtanen2020SciPy} and optionally \texttt{Matplotlib} for plotting \citep{Hunter2007Matplotlib}. All results in this paper are reproducible using the commands in Appendix~\ref{app:repro}.

\section{SIDM microphysics layer}
\subsection{Yukawa interaction model}
We consider elastic scattering of identical dark matter particles of mass $m_\chi$ mediated by a Yukawa potential with mediator mass $m_\phi$ and coupling $\alpha_\chi$,
\begin{equation}
V(r) = \pm \frac{\alpha_\chi}{r}\,e^{-m_\phi r},
\label{eq:yukawa}
\end{equation}
where the sign corresponds to attractive or repulsive interactions. This non-relativistic potential is a common effective description for a range of SIDM models \citep{TulinYu2018}.

\subsection{Transfer cross section}
Astrophysical observables are often more directly sensitive to the momentum-transfer cross section,
\begin{equation}
\sigma_T \equiv \int d\Omega\,(1-\cos\theta)\,\frac{d\sigma}{d\Omega},
\label{eq:sigmat_def}
\end{equation}
which suppresses forward scattering. (Some contexts use alternative angular weights for identical particles; \texttt{sidmkit} focuses on $\sigma_T$ as the default because it matches the convention of many astrophysical summaries \citep{Adhikari2022Review}.)

\subsection{Regimes and approximations}
Exact Yukawa scattering requires solving the Schr\"odinger equation, but standard approximations capture the main behaviour in most of parameter space \citep{TulinYu2018}. \texttt{sidmkit} implements:
\begin{itemize}
\item \textbf{Born regime:} valid for $\alpha_\chi m_\chi/m_\phi \ll 1$. The implementation follows the closed-form momentum-transfer expression used widely in the SIDM literature (see \texttt{sidmkit.cross\_sections} for the exact expression).
\item \textbf{Classical regime:} valid when the de Broglie wavelength is short compared to the interaction range. We use a standard piecewise approximation for $\sigma_T(\beta)$ where $\beta \propto \alpha_\chi m_\phi/(m_\chi v^2)$.
\item \textbf{Resonant / Hulth\'en approximation:} in parts of the non-perturbative regime, the Yukawa potential can be approximated by a Hulth\'en potential to capture resonant features.
\item \textbf{Partial-wave (numerical) option:} a direct phase-shift computation used for spot checks and debugging. It is \emph{not} recommended for large parameter scans without care, because it is slower and can fail in extreme resonant regions if the ODE solver requires prohibitively small step sizes.
\end{itemize}

A practical point: any ``auto'' regime selection is necessarily heuristic near regime boundaries. \texttt{sidmkit} therefore exposes the method choice explicitly and emits warnings when the dimensionless coupling $\alpha_\chi m_\chi/m_\phi$ is near unity.

\subsection{Velocity averaging}
Many astrophysical constraints depend on velocity moments such as $\langle\sigma_T v\rangle/m$. For isotropic Maxwellian one-particle velocities with 1D dispersion $\sigma_{1\mathrm{d}}$, the \emph{relative} speed distribution is Maxwellian with mean $\langle v_\mathrm{rel}\rangle = 4\sigma_{1\mathrm{d}}/\sqrt{\pi}$. We define the $n$-th moment
\begin{equation}
\left\langle \frac{\sigma_T v^n}{m_\chi} \right\rangle
= \int_0^\infty dv\, f_\mathrm{rel}(v;\sigma_{1\mathrm{d}})\,\frac{\sigma_T(v)}{m_\chi}\,v^n,
\label{eq:velavg}
\end{equation}
and evaluate it numerically using either Gauss--Laguerre quadrature (for integrals over $[0,\infty)$) or adaptive quadrature, with internal regression tests verifying agreement at the $10^{-9}$ level for representative cases.

\subsection{Scattering rate and an illustrative $r_1$ radius}
A commonly used halo-level diagnostic is the per-particle scattering rate,
\begin{equation}
\Gamma(r) = \frac{\rho(r)}{m_\chi}\,\left\langle \sigma_T v \right\rangle,
\label{eq:gamma}
\end{equation}
where $\rho(r)$ is the local dark matter density, and the velocity moment depends on the assumed velocity distribution. A simple core-formation proxy used in analytic SIDM treatments is the radius $r_1$ at which a typical particle has scattered once over a halo age $t_\mathrm{age}$ \citep{Kaplinghat2016Colliders},
\begin{equation}
\Gamma(r_1)\,t_\mathrm{age} \simeq 1.
\label{eq:r1}
\end{equation}
\texttt{sidmkit} includes an illustrative implementation for NFW halos to support order-of-magnitude estimates. We stress that this is \emph{not} a substitute for controlled SIDM simulations; it is best used for exploration and unit/regression checks.

\section{Rotation-curve fitting to SPARC \texttt{rotmod} files}
\subsection{SPARC \texttt{rotmod} format and baryonic decomposition}
SPARC provides rotation curves and baryonic decompositions for disk galaxies, including contributions from gas, stellar disk, and (when applicable) bulge \citep{Lelli2016SPARC}. In the \texttt{rotmod} format used here, each radial bin contains:
\[
\{r,\,V_\mathrm{obs},\,\sigma_V,\,V_\mathrm{gas},\,V_\mathrm{disk},\,V_\mathrm{bulge},\,\dots\}.
\]
The baryonic templates are velocities computed for a reference mass-to-light normalisation (effectively $\Upsilon_\star=1$ in the template convention). We model the total circular speed as
\begin{equation}
V_\mathrm{model}^2(r) = V_\mathrm{gas}^2(r)
+ \Upsilon_{\star,d}\,V_\mathrm{disk}^2(r)
+ \Upsilon_{\star,b}\,V_\mathrm{bulge}^2(r)
+ V_\mathrm{halo}^2(r;\bm{\theta}),
\label{eq:vmodel}
\end{equation}
where $\Upsilon_{\star,d}$ and $\Upsilon_{\star,b}$ are fitted disk/bulge mass-to-light factors and $\bm{\theta}$ are halo parameters.

\subsection{Halo models}
We fit two standard spherical halo profiles:

\paragraph{NFW.}
The Navarro--Frenk--White profile \citep{Navarro1997NFW} is
\begin{equation}
\rho_\mathrm{NFW}(r) = \frac{\rho_s}{(r/r_s)\,(1+r/r_s)^2}.
\end{equation}
The enclosed mass is
\begin{equation}
M_\mathrm{NFW}(r) = 4\pi\rho_s r_s^3\left[\ln(1+x) - \frac{x}{1+x}\right],\quad x\equiv r/r_s,
\end{equation}
and the circular speed is $V_\mathrm{halo}^2(r)=G M(r)/r$.

\paragraph{Burkert.}
The Burkert profile \citep{Burkert1995} is a cored phenomenological model,
\begin{equation}
\rho_\mathrm{Bur}(r) = \frac{\rho_0 r_0^3}{(r+r_0)\,(r^2+r_0^2)}.
\end{equation}
The analytic enclosed mass used in our implementation is
\begin{equation}
M_\mathrm{Bur}(r) = \pi\rho_0 r_0^3\left[\ln\!\left((1+x)^2(1+x^2)\right) - 2\arctan(x)\right],\quad x\equiv r/r_0.
\end{equation}

\subsection{Objective function, priors, and optimisation}
For each galaxy, we minimise a weighted least-squares objective equivalent to a Gaussian log-likelihood,
\begin{equation}
\chi^2(\bm{\theta}) = \sum_{i=1}^{N}\left(\frac{V_\mathrm{model}(r_i;\bm{\theta}) - V_{\mathrm{obs},i}}{\sigma_{V,i}}\right)^2.
\label{eq:chi2}
\end{equation}
We use bounded non-linear least squares (\texttt{scipy.optimize.least\_squares}) \citep{Virtanen2020SciPy} with broad but finite bounds on halo parameters and $\Upsilon_\star \ge 0$ to avoid pathological excursions. Optionally, we include weak Gaussian priors on $\Upsilon_{\star,d}$ and $\Upsilon_{\star,b}$ by appending corresponding residuals. Importantly, when we report $\chi^2$, AIC, and BIC, we compute them from the \emph{data-only} residuals at the best-fit point (priors excluded), so that information-criterion comparisons reflect the fit to rotation-curve data rather than the prior penalty.

We summarise fit quality using:
\begin{align}
\chi^2_\nu &= \chi^2/(N-k),\\
\mathrm{AIC} &= \chi^2 + 2k,\\
\mathrm{BIC} &= \chi^2 + k\ln N,
\label{eq:bic}
\end{align}
where $k$ is the number of fitted parameters. Since NFW and Burkert have the same $k$ in our setup (either 3 or 4 depending on bulge), $\Delta\mathrm{BIC}$ between them reduces to a constant multiple of $\Delta\chi^2$ for each galaxy. We use BIC only as a compact summary, not as a substitute for a full physical model comparison.

\subsection{Batch processing and chunking}
A submission-grade SPARC analysis must handle hundreds of galaxies reproducibly. To avoid long single runs and to support parallelisation, \texttt{sidmkit} implements chunked processing via \texttt{--skip} and \texttt{--limit}. Each chunk writes:
\begin{itemize}
\item per-galaxy JSON fit files,
\item a chunk-level \texttt{summary.csv} and \texttt{summary.json},
\item optional paper-style per-galaxy plots with residual panels.
\end{itemize}
Multiple chunks are merged into a single population summary, and a report command generates population figures and summary statistics.

\section{Validation and numerical checks}
\subsection{Internal benchmarks}
\texttt{sidmkit} includes a benchmark suite (\texttt{sidmkit benchmark}) that checks:
(i) scattering-rate unit consistency,
(ii) Maxwellian relative-speed normalisation and mean,
(iii) agreement between two independent velocity-averaging integrators (Gauss--Laguerre vs adaptive quadrature),
and (iv) optional slow checks comparing ``auto'' cross sections to a partial-wave calculation in representative regimes.

These checks are designed as regression tests: they help detect silent numerical drift when code is refactored. They do not replace validation against external published curves, which remains essential for any claim of astrophysical inference.

\subsection{SPARC fitter sanity checks}
For the dataset analysed here, all 191 \texttt{rotmod} files parsed cleanly (no missing columns after basic cleaning). We flag two reviewer-relevant diagnostics:
\begin{itemize}
\item \textbf{Parameter-bound saturation:} $21.5\%$ of NFW fits hit the upper bound on $\log_{10}(r_s/\mathrm{kpc})$ (set to $2.5$ in our default). This is a sign of degeneracy or model mismatch for those objects and can inflate $\Delta\mathrm{BIC}$ in favour of Burkert.
\item \textbf{Small-$N$ galaxies:} For 16 galaxies with $N\le k$, $\chi^2_\nu$ is not defined. We still report $\chi^2$ and information criteria, but reduced-$\chi^2$ summaries exclude these cases.
\end{itemize}
These diagnostics motivate future extensions (e.g., alternative parameterisations, hierarchical priors, distance/geometry uncertainties), but they already provide important context for interpreting population summaries.

\section{Results: batch fits for 191 \texttt{rotmod} galaxies}
\subsection{Dataset summary}
We processed 191 galaxies from the provided LTG+ETG \texttt{rotmod} bundles (SPARC) and performed two fits per galaxy (NFW and Burkert), totalling 382 fits. The number of rotation-curve points ranges from 2 to 115, with a median of 12. Approximately $25\%$ of galaxies include a non-trivial bulge template in the input files, in which case $\Upsilon_{\star,b}$ is fitted.

\subsection{Fit quality}
\begin{figure}[t]
\centering
\begin{subfigure}{0.49\linewidth}
\centering
\includegraphics[width=\linewidth]{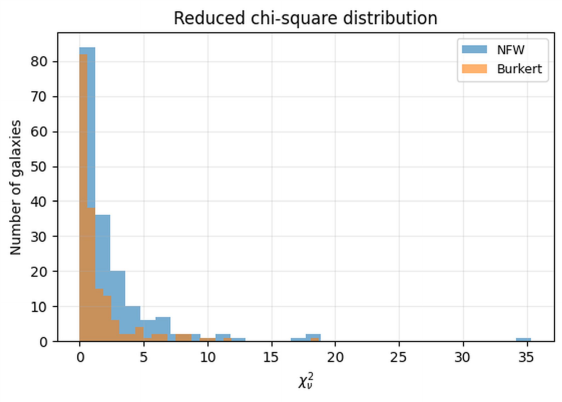}
\caption{Histogram of reduced $\chi^2$.}
\end{subfigure}
\hfill
\begin{subfigure}{0.49\linewidth}
\centering
\includegraphics[width=\linewidth]{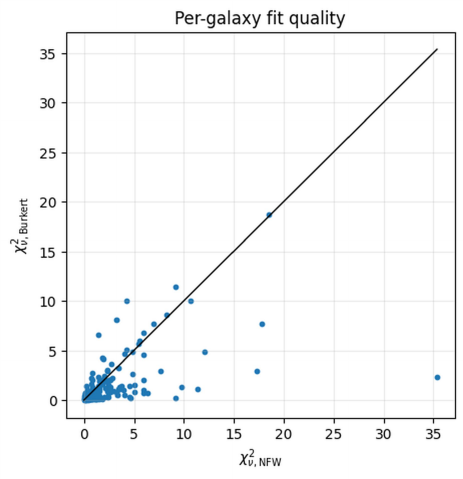}
\caption{$\chi^2_\nu$ scatter: Burkert vs NFW.}
\end{subfigure}
\caption{Fit quality across the sample. Reduced $\chi^2$ is shown only where $N>k$.}
\label{fig:chi2}
\end{figure}

Figure~\ref{fig:chi2} summarizes reduced $\chi^2$ distributions. For galaxies with positive degrees of freedom, the median values are:
\[
\tilde{\chi}^2_{\nu,\mathrm{NFW}} = 1.25,\qquad
\tilde{\chi}^2_{\nu,\mathrm{Bur}} = 0.71.
\]
The Burkert profile yields $\chi^2_\nu<1$ for $72.6\%$ of galaxies with defined $\chi^2_\nu$, compared to $44.6\%$ for NFW. This is consistent with the qualitative expectation that a cored phenomenology often tracks inner rotation curves more flexibly, though the physical origin of any ``core-like'' preference is not uniquely determined by these fits.

\subsection{Model comparison: $\Delta\mathrm{BIC}$}
\begin{figure}[t]
\centering
\begin{subfigure}{0.49\linewidth}
\centering
\includegraphics[width=\linewidth]{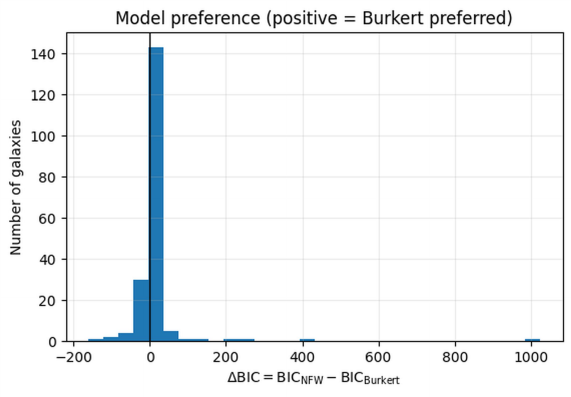}
\caption{Histogram of $\Delta\mathrm{BIC}$.}
\end{subfigure}
\hfill
\begin{subfigure}{0.49\linewidth}
\centering
\includegraphics[width=\linewidth]{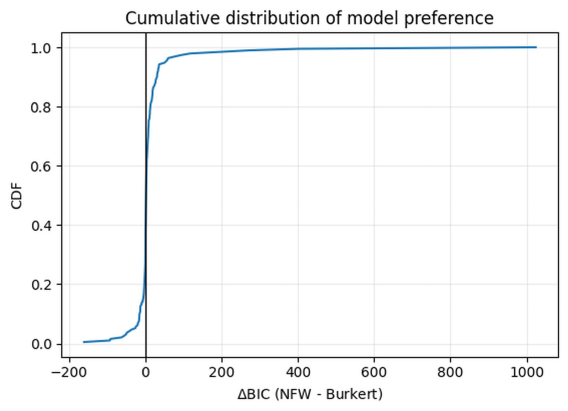}
\caption{CDF of $\Delta\mathrm{BIC}$.}
\end{subfigure}
\caption{Population-level model preference. Positive $\Delta\mathrm{BIC}$ indicates Burkert preference.}
\label{fig:dbic}
\end{figure}

We define
\begin{equation}
\Delta\mathrm{BIC} \equiv \mathrm{BIC}_\mathrm{NFW} - \mathrm{BIC}_\mathrm{Bur}.
\end{equation}
Positive values indicate a preference for Burkert. The distribution is shown in Figure~\ref{fig:dbic}. Key summary statistics are:
\begin{itemize}
\item $65.4\%$ of galaxies have $\Delta\mathrm{BIC}>0$ (Burkert preferred).
\item $32.5\%$ have $\Delta\mathrm{BIC}>6$ (commonly interpreted as ``strong'' preference).
\item $14.7\%$ have $\Delta\mathrm{BIC}<-6$ (strong NFW preference).
\item Median $\Delta\mathrm{BIC}=1.81$; mean $\Delta\mathrm{BIC}=12.9$ (heavy-tailed).
\end{itemize}
Because both models have the same parameter count in our setup, these numbers mainly reflect differences in $\chi^2$ rather than a complexity penalty.

\begin{figure}[t]
\centering
\includegraphics[width=0.7\linewidth]{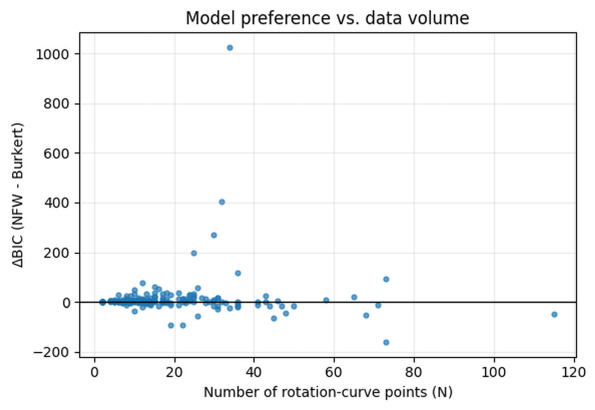}
\caption{$\Delta\mathrm{BIC}$ vs. number of rotation-curve points.}
\label{fig:dbic_n}
\end{figure}

We also tested whether $\Delta\mathrm{BIC}$ is trivially driven by the number of data points. Figure~\ref{fig:dbic_n} shows only a weak linear correlation between $\Delta\mathrm{BIC}$ and $N$, suggesting that preference is not merely an artefact of sample size, although large-$N$ galaxies do produce more extreme values when the fit mismatch is systematic.

\subsection{Fitted parameter distributions and degeneracies}
\begin{figure}[t]
\centering
\begin{subfigure}{0.49\linewidth}
\centering
\includegraphics[width=\linewidth]{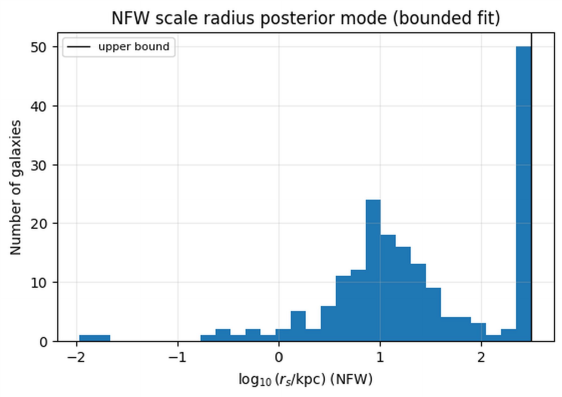}
\caption{NFW scale radius.}
\end{subfigure}
\hfill
\begin{subfigure}{0.49\linewidth}
\centering
\includegraphics[width=\linewidth]{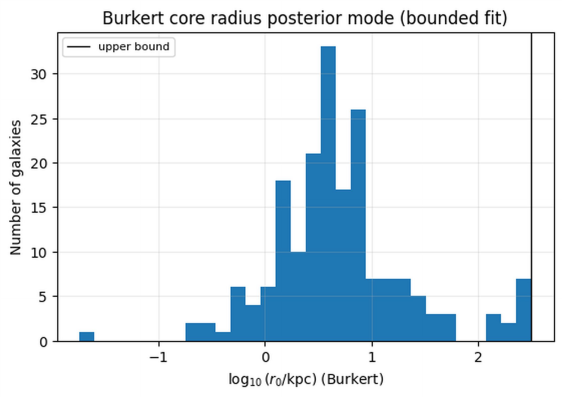}
\caption{Burkert core radius.}
\end{subfigure}

\vspace{0.5em}
\begin{subfigure}{0.49\linewidth}
\centering
\includegraphics[width=\linewidth]{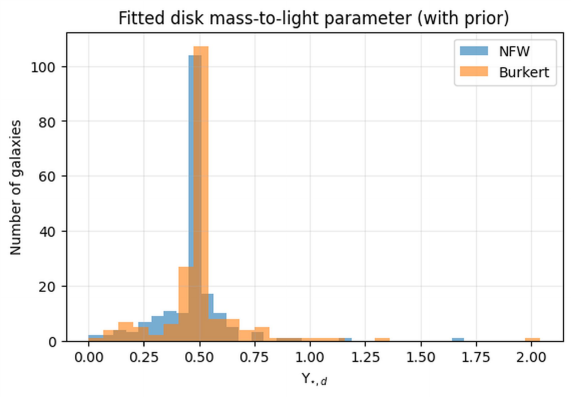}
\caption{Disk mass-to-light parameter.}
\end{subfigure}
\hfill
\begin{subfigure}{0.49\linewidth}
\centering
\includegraphics[width=\linewidth]{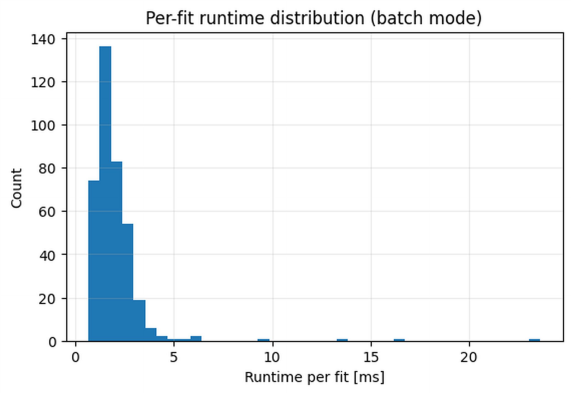}
\caption{Runtime per fit (batch mode).}
\end{subfigure}
\caption{Distributions of fitted parameters and computational performance.}
\label{fig:params}
\end{figure}

Figure~\ref{fig:params} shows distributions of fitted halo scale radii and the disk mass-to-light parameter. Two important observations:
\begin{enumerate}
\item A visible pile-up of NFW fits at the upper bound of $\log_{10}(r_s/\mathrm{kpc})=2.5$ indicates that, for a non-negligible fraction of galaxies, the chosen parameterisation/bounds interact with the optimisation. This should be treated as a red flag for any downstream physical inference.
\item The fitted $\Upsilon_{\star,d}$ distribution is narrow because we include a weak Gaussian prior by default. Without priors (\texttt{--no-priors}), the distribution broadens and can absorb some inner-curve mismatch. For that reason, we do not interpret fitted $\Upsilon_\star$ values as astrophysical measurements here.
\end{enumerate}

\subsection{Representative galaxy fits}
\begin{figure}[t]
\centering
\begin{subfigure}{0.49\linewidth}
\centering
\includegraphics[width=\linewidth]{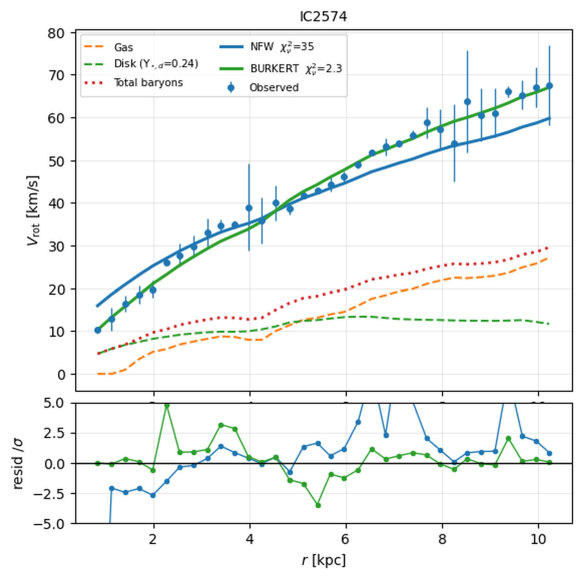}
\caption{IC2574 (strong Burkert preference).}
\end{subfigure}
\hfill
\begin{subfigure}{0.49\linewidth}
\centering
\includegraphics[width=\linewidth]{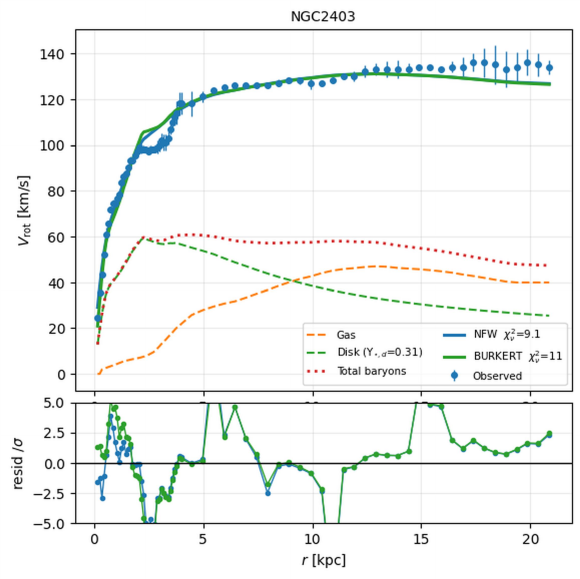}
\caption{NGC2403 (strong NFW preference).}
\end{subfigure}

\vspace{0.5em}
\begin{subfigure}{0.49\linewidth}
\centering
\includegraphics[width=\linewidth]{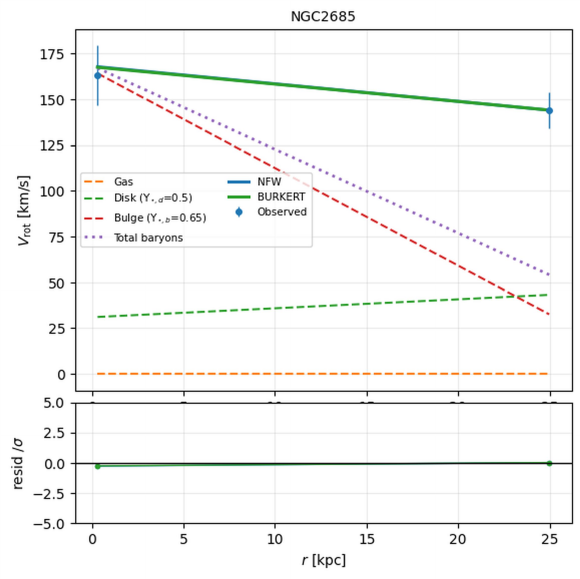}
\caption{NGC2685 (moderate preference).}
\end{subfigure}
\hfill
\begin{subfigure}{0.49\linewidth}
\centering
\includegraphics[width=\linewidth]{figures/delta_bic_hist.pdf}
\caption{Population context.}
\end{subfigure}
\caption{Example fits and population context.}
\label{fig:examples}
\end{figure}

Figure~\ref{fig:examples} shows representative per-galaxy fits with residual panels. The full set of per-galaxy plots (when generated) is intended as supplementary material rather than being reproduced in the main text.

\subsection{Best/worst cases (diagnostic table)}
Table~\ref{tab:extremes} lists a few galaxies with the largest absolute $\Delta\mathrm{BIC}$. These are useful diagnostics: extreme preferences can reflect genuine shape differences, but they can also indicate optimiser-bound interactions (especially for NFW) or baryonic-systematic mismatches.

\begin{table}[t]
\centering
\caption{Examples of extreme model preference (data-only $\Delta\mathrm{BIC}$).}
\label{tab:extremes}
\begin{tabular}{lrrrr}
\toprule
Galaxy & $\Delta\mathrm{BIC}$ & $N$ & Distance [Mpc] & Preferred \\
\midrule
IC2574 & 1024.2 & 34 & 3.91 & Burkert \\
IC4202 & 402.0 & 32 & 100.4 & Burkert \\
ESO563-G021 & 271.5 & 30 & 60.8 & Burkert \\
NGC3109 & 197.0 & 25 & 1.33 & Burkert \\
UGC11455 & 117.9 & 36 & 78.6 & Burkert \\
\midrule
NGC2403 & $-160.4$ & 73 & 3.16 & NFW \\
UGC00128 & $-92.7$ & 22 & 64.5 & NFW \\
NGC5371 & $-91.1$ & 19 & 39.7 & NFW \\
UGC06786 & $-61.8$ & 45 & 29.3 & NFW \\
NGC0247 & $-55.3$ & 26 & 3.70 & NFW \\
\bottomrule
\end{tabular}
\end{table}

\section{Discussion: how SPARC fits connect (and do not connect) to SIDM}
Rotation curves are one of the clearest probes of inner halo structure, and SPARC provides an unusually uniform set of high-quality curves and baryonic decompositions \citep{Lelli2016SPARC}. This makes SPARC valuable for SIDM in two complementary ways:
\begin{enumerate}
\item \textbf{Phenomenological calibration.} Even without a microphysical mapping, cored profiles (Burkert-like) often provide better empirical fits to inner rotation curves. This can be used to identify which objects are ``core-sensitive'' and to summarise core-radius distributions.
\item \textbf{A bridge layer.} SIDM microphysics predicts velocity-dependent $\sigma/m$; halos of different characteristic velocities probe different parts of $\sigma(v)$. In principle, one can connect SPARC-inferred inner structure (e.g., preferred core sizes) to SIDM parameters using an analytic model or simulation-calibrated mapping \citep{Kaplinghat2016Colliders, TulinYu2018}.
\end{enumerate}

However, our current SPARC pipeline deliberately does \emph{not} claim such a mapping. Reasons include:
\begin{itemize}
\item \textbf{Baryonic and geometric systematics:} uncertainties in inclination, distance, non-circular motions, and mass-to-light ratios can materially change inner-curve interpretation.
\item \textbf{Model incompleteness:} NFW and Burkert are simplified spherical models; real halos may be triaxial, contracted/expanded by baryons, or better described by other profiles.
\item \textbf{Inference structure:} a convincing SIDM inference from SPARC would likely require a hierarchical model across galaxies (or at least robust marginalisation over nuisance parameters), not independent per-galaxy point estimates.
\end{itemize}
We therefore interpret the results here as a \emph{reproducible baseline} that can be extended toward SIDM inference, rather than as an end state.

\section{Conclusions}
We presented \texttt{sidmkit}, an open and reproducible Python toolkit for SIDM phenomenology and batch SPARC rotation-curve fitting. The package provides:
(i) Yukawa-model $\sigma_T(v)$ computations with standard approximations and optional partial-wave checks,
(ii) Maxwellian velocity-moment averaging utilities,
(iii) curated literature \emph{summary} constraints for rapid scans and regression tests,
and (iv) a submission-grade SPARC \texttt{rotmod} batch fitting pipeline with chunking, merged summaries, and paper-style plots.

Applied to 191 \texttt{rotmod} galaxies, Burkert profiles outperform NFW for a majority of objects in a simple $\chi^2$/BIC sense. The analysis also surfaces important diagnostics (e.g., NFW scale-radius bound saturation) that should be addressed before attempting any high-stakes microphysical inference.

\section*{Data and software availability}
The \texttt{sidmkit} source, scripts, and the SPARC batch outputs supporting the figures and tables in this paper are included in the accompanying submission archive. SPARC rotation-curve data are publicly available \citep{Lelli2016SPARC}. All plots in this manuscript can be regenerated using the commands in Appendix~\ref{app:repro}.

%\section*{Software Availability}
The \texttt{sidmkit} toolkit is released as open-source software and is publicly available.
The core package is distributed via the Python Package Index (PyPI) at\\
\url{https://pypi.org/project/sidmkit/}.\\
The source code and development repository are hosted on GitHub at\\
\url{https://github.com/nalin-dhiman/sidmkit},\\
with the associated batch analysis and SPARC pipeline available at\\
\url{https://github.com/nalin-dhiman/SIDMkit_pipeline}.\\
The codebase is structured to facilitate reproducible analyses and extension by external users.

\appendix
\section{Reproducibility commands}
\label{app:repro}
Below is a minimal, end-to-end workflow that reproduces the SPARC batch results and population report.

\paragraph{1) Install.}
\begin{verbatim}
python -m pip install -U pip
pip install -e .
\end{verbatim}

\paragraph{2) Run chunked fits (example).}
\begin{verbatim}
python -m sidmkit.sparc_batch batch \
  --inputs src/sidmkit/sparc_data/Rotmod_LTG src/sidmkit/sparc_data/Rotmod_ETG \
  --outdir outputs/sparc_chunks/chunk_0 \
  --skip 0 --limit 25 \
  --plots --plot-format pdf
\end{verbatim}
Repeat with \texttt{--skip 25}, \texttt{--skip 50}, \dots\ to cover the full dataset, optionally in parallel.

\paragraph{3) Merge chunk summaries.}
\begin{verbatim}
python -m sidmkit.sparc_batch merge \
  --inputs outputs/sparc_chunks/chunk_*/summary.json \
  --out outputs/sparc_all_summary.json
\end{verbatim}

\paragraph{4) Generate population report.}
\begin{verbatim}
python -m sidmkit.sparc_batch report \
  --summary-json outputs/sparc_all_summary.json \
  --outdir outputs/sparc_report
\end{verbatim}

\end{document}